\documentclass[pra,amsmath,aps,twocolumn,superscriptaddress,letterpaper,showpacs]{revtex4}
\usepackage{graphicx,xspace,units,subfigure}
\usepackage{float}
\usepackage{amsmath}
\usepackage{amssymb}
\usepackage[normalem]{ulem}
\usepackage{color}
\newcommand{\Jr}{J_R}
\newcommand{\Jb}{J_B}
\newcommand{\Jj}{J_J}
\newcommand{\epsj}{\epsilon_J}
\newcommand{\epsr}{\epsilon_R}
\newcommand{\wj}{w_J}
\newcommand{\wrod}{w_R}
\newcommand{\vecb}{\vec{b}}
\newcommand{\vecc}{\vec{b}'}
\newcommand{\vecr}{\vec{r}}
\newcommand{\hatb}{\hat{b}}
\newcommand{\hatc}{\hat{b}'}
\newcommand{\hatr}{\hat{r}}
\newcommand{\R}{R}
\newcommand{\B}{B}
\newcommand{\GR}{\Gamma_\R}
\newcommand{\GB}{\Gamma_\B}

\begin{document}

\title{Melting of persistent double--stranded polymers}
\author{Sahand Jamal Rahi}
\email{sjrahi@mit.edu}
\author{Mark Peter Hertzberg}
\author{Mehran Kardar}
\affiliation{Massachusetts Institute of Technology, Department of
  Physics, 77 Massachusetts Avenue, Cambridge, MA 02139, USA}

\begin{abstract}   
Motivated by recent DNA-pulling experiments, we revisit the
Poland-Scheraga model of melting a double-stranded polymer.  We
include distinct bending rigidities for both the double-stranded
segments, and the single-stranded segments forming a bubble.  There is
also bending stiffness at the branch points between the two segment
types.  The transfer matrix technique for single persistent chains is
generalized to describe the branching bubbles.  Properties of
spherical harmonics are then exploited in truncating and numerically
solving the resulting transfer matrix.  This allows efficient
computation of phase diagrams and force-extension curves (isotherms). While the main focus is on exposition of the transfer matrix technique, we provide general arguments for a reentrant melting transition in stiff double strands.
Our theoretical approach can also be
extended to study polymers with bubbles of any number of strands,
with potential applications to molecules such as collagen.
\end{abstract}


\maketitle

\section{Introduction}
\label{sec:Introduction}
Single-molecule micromanipulation techniques have opened up new
opportunities for measurements and studies of polymers.  Smith et al.
pioneered~\cite{Smith92} stretching experiments of double-stranded DNA
(dsDNA) and, along with others, observed that at high forces of about
$65\unit{pN}$, DNA extends to $1.7$ times its contour
length\cite{Bensimon95,Smith96,Cluzel96,Rief99,Clausen-Schaumann00}.
These investigators believe that the stretching transforms B-DNA,
which is DNA in its natural state, to a new, extended state, named
S-DNA. Modeling studies and simulations were carried out to
characterize this putative new state of
DNA.\cite{Konrad96,Kosikov99,Lebrun96} Subsequently, Storm and
Nelson~\cite{Storm03} proposed a statistical model of DNA as a discrete
persistent chain (DPC) with two monomer flavors of different lengths
and stiffnesses, and fit their parameters successfully to experimental
data. However, Williams, Rouzina, Bloomfield, and co-workers have
argued on the basis of their own experiments that S-DNA is not a
new state of the molecule, but merely DNA that is melted to two single-stranded
DNA (ssDNA)
fragments.\cite{Williams02,Williams01-1,Williams01-2,Rouzina01-1,Rouzina01-2,Vladescu05}
Furthermore, they deem the aforementioned modeling and simulations of
S-DNA as contradicting experimental data. Furthering this controversy,
Cocco et al.\cite{Cocco04} reexamine the experimental data and argue
in favor of S-DNA, Whitelam et al.\cite{Whitelam06} do so based on
kinetics, while Piana~\cite{Piana05} observes melting in simulations of
short stretches of DNA.

In 1966 Poland and Scheraga\cite{Poland66} introduced a simple statistical model
for  the melting of the dsDNA to two ssDNA fragments, which has proved quite illuminating.
In this model, configurations of partially melted DNA are represented by
alternating segments of dsDNA, and denatured pairs of single strands forming `bubbles.'  
To make the model analytically tractable,  certain features of DNA such as
excluded volume, bending rigidity, and sequence inhomogeneity are typically left out.  
With the later inclusion of excluded volume effects, the model is well
suited for characterizing the nature of the melting transition, and
its universality.  For comprehensive (but older) reviews see
Refs.~\cite{Wartell85,Gotoh81}; some newer results are described
in, e.g. Ref.~\cite{Kafri02}.  More recently, the phase diagram of the
model has been studied in the presence of a stretching force~\cite{Hanke08, Rudnick07}.  
This is important, since even the
experiments disputing the formation of S-DNA at $65\unit{pN}$ do
observe melting induced stretching at other
forces\cite{Rief99,Clausen-Schaumann00}.  The effect of bending
rigidity is still left out in the newer studies, making comparisons to
experiment questionable.  The aim of this paper is to facilitate the
ongoing debate by providing a model that accounts for the bending
rigidity of the polymer (while leaving out excluded volume effects). 

While we hope that our results and phase diagrams provide an
additional perspective into this system, our main accomplishment is the
extension of the transfer matrix method used for a single persistent polymer
(worm-like chain) to the melting of a double-stranded polymer.
The remainder of the paper is an exposition of our method, and is organized
as follows.
The generalized Poland--Scheraga model with three types of bending rigidity
is introduced in Sec.~\ref{sec: Energetics}, and the corresponding three contributions
to transfer matrices are developed in Sec.~\ref{sec:TM}.
As described in Sec.~\ref{sec:Results}, numerical results can be obtained by truncating the resulting
transfer matrices in a basis of spherical harmonics. In particular, we provide phase diagrams
(in force and temperature) and force--extension curves, along with the native (double stranded)
fraction. We augment numerical results with physical explanations of the observed trends. In particular, we provide a rather general characterization of the slop of the phase boundary which explains the potential reentrant character of force induced melting.
Various technical details of the calculation are relegated to the Appendices. 

\section{Model}
\label{sec:Model}
\subsection{Energetics}\label{sec: Energetics}
\begin{figure}[h]
\centering
\includegraphics[width=8.5cm]{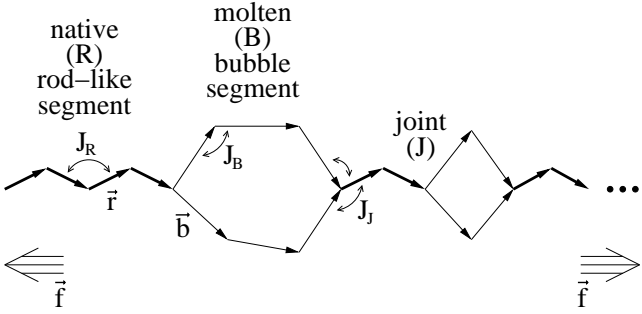}
\caption{A typical polymer configuration of our model, as depicted here,
consists of segments \R, solid arrows, which we imagine to be dsDNA,
alternating with `bubbles' \B~ made of two strands of ssDNA (light
arrows).  The two segment types have unit (monomer) lengths $|\vecr|$
or $|\vecb|$, and bending costs of $\Jr$ or $\Jb$, respectively.
There is an additional bending constant $\Jj$, at the branching
points, and a weight $\wj=e^{\epsj}$ for each joint.  The
energetic advantage (binding energy) of the \R~segments is represented
by a weight $\wrod=e^{\epsr}$ per step.}
\label{fig:arrows}
\end{figure}

As illustrated in Fig.~\ref{fig:arrows}, a typical configuration of our model
polymer consists of an alternating sequence of native segments \R, and locally molten
pairs of strands forming a bubble \B.
Successive segments are indexed by $i$, and contain $N_{\R,i}$ or $2 N_{\B,i}$
monomers, respectively. 
In the original Poland-Scheraga model~\cite{Poland66}, the 
\R~segments were treated as stiff 'r'ods. 
We treat these segments as semi-flexible chains, such that the energy
of a segment of $N_{\R,i}$ monomers is given by
\begin{equation}
\begin{split}\label{eq:Erods}
- \beta E_{\R} & = \sum_{j=1}^{N_{\R}-1}
\left( \Jr ~\hatr_{j} \cdot \hatr_{j+1} + \vecr_{j} \cdot \vec{f}\right) \\
               & + \vecr_{N_{\R}} \cdot \vec{f} + N_{\R} \epsr \quad.
\end{split}
\end{equation}
Here, $\vecr_{j}=|\vecr|\hatr_{j}$ is the displacement of the $j$'th
`monomer' of the segment, all of which have equal length, but may
point in any direction.  The coupling $\Jr$ parameterizes the cost of
bending neighboring monomers.  The force $\vec{f}$ stretches the
polymer, and $\epsr$ is an additional contribution to the energy
difference between a native \R~unit compared to the molted strands of
\B~units. 
Note that (for each configuration, and discounting bending
costs) the net energy difference between bound and unbound segments
(the binding energy) is $k_BT(\Jr+\epsr)$ per base-pair.  (For ease of
notation, the index $i$ denoting the $i$'th \R~segment has been
dropped from all variables above.)

Similarly, the energy of a molten \B~region, described by $2 N_{\B,i}$
units $\vecb_{j}$ and $\vecc_{j}$ (for the two strands) is given by
\begin{equation}
\begin{split}
- \beta E'_{\B} & = \sum_{j=1}^{N_{\B}-1}
\left( \Jb~ \hatb_{j} \cdot \hatb_{j+1} + \vecb_{j} \cdot \frac{\vec{f}}{2}\right) \\
              & + \sum_{j=1}^{N_{\B}-1}
\left( \Jb~ \hatc_{j} \cdot \hatc_{j+1} + \vecc_{j} \cdot \frac{\vec{f}}{2}\right) \\
              & + \vecb_{N_{\B}} \cdot \frac{\vec{f}}{2} + \vecc_{N_{\B}} \cdot \frac{\vec{f}}{2}\quad.
\end{split}
\end{equation}
Again, the implicit index $i$ numbering the $i$'th \B~segment has been
omitted.  The allowed configurations are constrained by
$\vec{R}_\B=\sum_{j=1}^{N_{\B}} \vecb_{j} = \sum_{j=1}^{N_{\B}}
\vecc_{j}$, to ensure that the two branches of the bubble end at the same
point.  It is indeed this constraint (emphasized by the primed $E_{\B}$) that
allows distributing the energy cost of stretching by the force
$\vec{f}$ symmetrically between the two branches.

Finally, there is a joint when the $N_{\R,i}$'th (last) element of the $i$'th \R~segment 
branches into the first elements of the $i$'th \B~segment, to which
we associate an energy
\begin{equation}
- \beta E_{J,\R\B} =
\Jj~ \hatr_{N_{\R}} \cdot \hatb_{1} +
\Jj~ \hatr_{N_{\R}} \cdot \hatc_{1} +
\epsj\quad.
\end{equation}
Similarly at the point where the $i$'th \B~segment meets the $(i+1)$'th \R~segment,
the energy is
\begin{equation}
- \beta E_{J,\B\R} =
\Jj~ \hatb_{N_{\B}} \cdot \hatr_{1} +
\Jj~ \hatc_{N_{\B}} \cdot \hatr_{1} +
\epsj\quad.
\end{equation}
The overall energy of $M$ alternating \R-\B~segments of sizes $\{N_{\R,i},N_{\B,i}\}$ is thus
\begin{equation}
\begin{split}
&\beta E'\left[N_{\R,1},N_{\B,1},N_{\R,2},\cdots,N_{\B,M}\right] = \\
&\sum_{i=1}^M \beta E_{\R,i} + \beta E_{J,\R\B,i} + \beta E'_{\B,i} +
\beta E_{J,\B\R,i}\quad.
\end{split}
\label{eq:Etot}
\end{equation}
(The above formula applies to configurations which start with an \R~segment 
and end with a \B~segment. We expect the results for long polymers to be 
independent of the choice of boundary conditions.)

Computations are most easily performed in a grand canonical ensemble in
which we sum over all possible polymer lengths, with a chemical potential
$\mu/\beta$ per monomer.
The grand partition function is then calculated from
\begin{equation}
\Gamma = \int_{\mathbf{S^2}}' \sum_{N} e^{N \mu}
\sum_{\{N_{\R,i},N_{\B,i}\}_{i=1}^M} e^{- \beta E'\left[N_{\R,1},\cdots,N_{\B,M}\right]}\quad,
\label{eq:Gamma1}
\end{equation}
where $N=\sum_{i=1}^M N_{\R,i} + N_{\B,i}$ is the native polymer length.
The integrations are over all directions of the monomer
vectors $\hatr$, $\hatb$, and $\hatc$, provided that the bubble--closing
constraints are satisfied. This can be ensured by inserting
$\delta$-functions for each bubble segment, as
\begin{equation}
\delta\left(\sum_{j=1}^{N_{\B,i}} \vecb_{j} - \sum_{j=1}^{N_{\B,i}} \vecc_{j}\right)
= \int \,\frac{d^3 \vec{k}}{(2\pi)^3}\, e^{i (\sum \vecb_{j} -
\sum \vecc_{j}) \cdot \vec{k}}
\quad .
\end{equation}

\subsection{Transfer Matrix Formulation}\label{sec:TM}
The one-dimensional character of the energy in Eq.~(\ref{eq:Etot})
suggests a transfer matrix approach to the problem.  This is indeed a
standard tool for the study of semi-flexible chains
\cite{Kramers41,Storm03,Wiggins05,Yan03,Yan05}, but requires additional
elaboration to treat the bubbles.  
Below, we shall develop step by step
the contributions from the two segment types, and the joints in
between, to the overall transfer matrix.

\subsubsection{\R~segments}
The Boltzmann weight in Eq.~(\ref{eq:Gamma1}) involves a product of exponentials,
similar in form to plane waves. Such exponentials can be expanded in a basis of 
spherical harmonics and Bessel functions, which then allows the integrations over 
the orientations $\hatr$, $\hatb$, and $\hatc$.
For example, integrating over the unit vector $\hatr_{n}$ of an \R~segment yields
\begin{equation}
\begin{split}
\int_{\mathbf{S^2}} e^{\cdots +
\Jr \hatr_{n-1} \cdot \hatr_{n} +
\Jr \hatr_{n} \cdot \hatr_{n+1} +
\vecr_n \cdot \vec{f} +
\epsr + \mu + \cdots}
\,d^2 \hatr_n \\
= [\cdots Y_\alpha^{*}(\hatr_{n-1})]
\left(T_\R\right)_{\alpha,\beta}
[Y_\beta(\hatr_{n+1}) \cdots]\, ,
\end{split}
\label{eq:expandD}
\end{equation}
where summation over repeated indices is implied. 
Greek letters stand for
elements of the angular momentum basis $|l, m\rangle$, e.g., $\alpha$
stands for $(l_\alpha, m_\alpha)$, and the transfer matrix elements are
\begin{equation}
\left(T_\R\right)_{\alpha,\beta} = (4\pi)^2 C_{\alpha,\bar{\beta},\gamma}
i_\beta(\Jr) i_\gamma(|\vec{f}| |\vecr|) 
Y_\gamma^{*}(\hat{f})~\wrod~z~.
\label{eq:TR}
\end{equation}
Here, $i_\alpha$ is the modified spherical Bessel function of the
first kind of order $l_\alpha$;
$C_{\alpha,\bar{\beta},\gamma}\equiv\int_{\mathbf{S^2}}
Y_\alpha(\hatr) Y_\beta^{*}(\hatr) Y_\gamma(\hatr) \, d^2\hatr$ is
closely related to tabulated Gaunt coefficients, which can be expressed
in terms of Wigner $3j$-symbols, see Appendix~\ref{app:Gaunt}; and each
unit of an \R~segment carrier a fugacity $z=e^\mu$, and the binding
weight $\wrod=e^{\epsr}$ defined earlier. To make the notation uniform
and simple, a bar placed over an index of $C$,
e.g. $C_{\alpha,\bar{\beta},\gamma}$, indicates that the corresponding
spherical harmonic under the integral shall be complex conjugated. The
repeated $\gamma$ index implies a (finite) sum.  The expression
simplifies if the force $\vec{f}$ is chosen to point along the
$\hat{z}$ direction, in which case $(T_R)_{\alpha,\beta} \propto
\delta_{m_\alpha,m_\beta}$.
Note that the transfer matrix is asymmetric, as we have included $i_\beta(\Jr)$, 
but not $i_\alpha(\Jr)$ to avoid double-counting.

\subsubsection{\B~segments}
A similar computation for the two bubble strands yields two transfer
matrices. These must be combined into one matrix to be usable in the
later steps. Thus, the basis elements $|l,m\rangle$, $|l',m'\rangle$
are combined into one product basis element $|l,m\rangle \otimes
|l',m'\rangle$ with one-letter abbreviation
$\tilde{\alpha}\equiv(\alpha,\alpha')=((l_\alpha,m_\alpha),(l_{\alpha'},m_{\alpha'}))$, and
\begin{equation}
\begin{split}
\iint_{\mathbf{S^2}} \frac{d^3\vec{k}}{(2 \pi)^3}
\begin{array}{l}
d^2 \hatb_n \\
d^2 \hatc_n
\end{array}
\begin{array}{l}
e^{\cdots +
\Jb \hatb_{n-1} \cdot \hatb_{n} +
\Jb \hatb_{n} \cdot \hatb_{n+1} +
\vecb_n \cdot \frac{\vec{f}}{2} +
i\vecb_n \cdot \vec{k} +
\cdots} \\
e^{\cdots +
\Jb \hatc_{n-1} \cdot \hatc_{n} +
\Jb \hatc_{n} \cdot \hatc_{n+1} +
\vecc_n \cdot \frac{\vec{f}}{2} -
i\vecc_n \cdot \vec{k} +
\cdots}
\end{array}
z
 \\
=
\int \frac{d^3 \vec{k}}{(2 \pi)^3}
\begin{array}{l}
\left[\cdots Y_\alpha^{*}(\hatb_{n-1})\times\right] \\
\left[\cdots Y_{\alpha'}^{*}(\hatc_{n-1})\times\right]
\end{array}
(T_\B(\vec{k}))_{\tilde{\alpha},\tilde{\beta}}
\begin{array}{l}
\left[\times Y_\beta(\hatb_{n+1}) \cdots\right] \\
\left[\times Y_{\beta'}(\hatc_{n+1}) \cdots\right]
\end{array}
\\
\end{split} 
\,, 
\label{eq:expandB}
\end{equation}
where
\begin{equation}
\begin{split}
(T_\B(\vec{k}))_{\tilde{\alpha},\tilde{\beta}} = 
 z \int
Y_\alpha(\hatb)
e^{\vecb \cdot \frac{\vec{f}}{2} +
i\vecb \cdot \vec{k}}
Y_\beta^{*}(\hatb) (4\pi) i_\beta(\Jb)
\,d^2\hatb \\
\times \int
Y_{\alpha'}(\hatc)
e^{\vecc \cdot \frac{\vec{f}}{2} -
i\vecc \cdot \vec{k}}
Y_{\beta'}^{*}(\hatc) (4\pi) i_{\beta'}(\Jb)
\,d^2\hatc \quad.\\
\end{split}
\label{eq:TB}
\end{equation}

This transfer matrix is, in general, very big. If the spherical
harmonic basis elements are cut off at some $l_{\text{max}}$ for
numerical evaluation, there are $(l_{\text{max}}+1)^2$ basis elements
to consider since $m\in\{-l,\cdots,l\}$ for each $0\leq l\leq l_{\text{max}}$. 
This means that there are
$((l_{\text{max}}+1)^2)^2$ \emph{product} basis elements, which is the
number  of rows or columns of the transfer matrix $T_\B(\vec{k})$! With a little
trick this big matrix can be reduced to block diagonal form with the
biggest submatrix having size $(l_{\text{max}}+1)^2$:

\noindent Consider the first integral in Eq.~(\ref{eq:TB}) and let
$\vec{v}=i\vec{k}+{\vec{f}}/{2}$ indicate the vector in the
exponent. One can rotate the complex vector $\vec{v}$ into the
$\hat{z}$-direction, such that
\begin{equation}
\begin{split}
&(4\pi) \int
Y_\alpha(\hatb)
e^{\vecb \cdot \vec{v}}
Y_\beta^{*}(\hatb)
i_\beta(\Jb)
\,d^2\hatb = \\
&(4\pi) \, \mathcal{D}_{\alpha,\mu}
\left(
\int
Y_\mu(\hatb)
e^{\vecb \cdot |\vec{v}| \hat{z}}
Y_\nu^{*}(\hatb)
i_\nu(\Jb)
\,d^2\hatb
\right)
\mathcal{D}_{\nu,\beta}^{-1} = \\
&(4\pi)^2 \mathcal{D}_{\alpha,\mu}
\left(
C_{\mu,\bar{\nu},\gamma}
i_\nu(\Jb)
i_\gamma(|\vec{v}||\vecb|)
Y_\gamma^{*}(\hat{z})
\right)
\mathcal{D}_{\nu,\beta}^{-1}~, \\
\end{split}
\end{equation}
where $\mathcal{D}(\phi,\theta,0)$ is the rotation matrix (Wigner-D matrix
or Wigner-D function) for quantum mechanical angular momentum
states. The spherical coordinate angles $\theta$ (from the $z$-axis) and
$\phi$ (from the $x$-axis) are the angles by which the
$\hat{z}$-direction rotates into the $\vec{v}$-direction. The rotation
matrices are usually parametrized by Euler angles. The first Euler
angle corresponds to $\phi$, the second to $\theta$, and the third to
zero. Since $\vec{v}$ is complex, the angles of rotation are complex
and $|\vec{v}|$, which is the usual Euclidean norm of $\vec{v}$, is
also complex, as $|\vec{v}|^2\equiv \vec{v}\cdot\vec{v} =f^2/4-k^2+i \vec{k}\cdot\vec{f}$.
Note that $i_\beta(\Jb)$ in the first line can be replaced by
$i_\nu(\Jb)$ because $i_\beta$ only depends on $l_\beta$ while
$\mathcal{D}$ only mixes states with the same total angular momenta, i.e.
$\mathcal{D}_{\nu,\beta} \propto \delta_{l_\nu,l_\beta}$. Although the
dependence on $\vec{k}$ is now both in $\mathcal{D}$ as well as in
$i_\gamma$, the problem is computationally much simpler since
$C_{\mu,\bar{\nu},\gamma} Y_\gamma^{*}(\hat{z}) \propto
\delta_{m_\mu,m_\nu}$.

\subsubsection{Joints}
Similar manipulations lead to transfer matrices at the branching points of
\begin{equation}
\begin{split}
(T_{J,\R \B})_{\alpha,\tilde{\beta}} & = (4\pi)^3
C_{\alpha,\gamma,\bar{\xi}} C_{\bar{\beta},\bar{\beta}',\xi}
i_\gamma(|\vec{f}| |\vecr|) Y_\gamma^{*}(\hat{f}) \\
& \times i_\beta(\Jj) i_{\beta'}(\Jj) ~\wrod ~\wj~ z~,
\end{split}
\end{equation}
and
\begin{equation}
\begin{split}
(T_{J,\B \R})_{\tilde{\alpha},\beta} & = (4 \pi)^2
C_{\bar{\alpha},\bar{\alpha}',\xi} C_{\beta,\gamma,\bar{\xi}}
i_\gamma(|\vec{f}| |\vecr|) Y_\gamma^{*}(\hat{f}) \\
& \times \frac{i_\alpha(\Jj) i_{\alpha'}(\Jj)}
{i_\alpha(\Jb) i_{\alpha'}(\Jb)} i_\beta(\Jr) ~\wrod ~\wj ~z~.
\end{split}
\end{equation}
Once more the lack of symmetry between the two cases reflects our choice of
including the bending energy from the next (but not the previous) segment
in the transfer matrix. 
The absence of the joint energy $\Jj$ greatly simplifies the problem,
as the \R~ and \B~ segments can then be treated independently. This
happens because $i_\alpha(0) = 0$ unless $l_\alpha=0$.

Combining these expressions, we can express the partition function
$\Gamma$ in Eq.~(\ref{eq:Gamma1}) in terms of the transfer matrices
(the irrelevant prefactors have been omitted), as
\begin{equation}
\begin{split}
\Gamma \propto
&
\left.
\sum_{M=1}^{\infty}
\left[
\left(
\sum_{i=0}^\infty T_\R^i
\right)
T_{J,\R \B}
\left(
\int \frac{d^3 \vec{k}}{(2 \pi)^3}
\sum_{i=2}^\infty
T_\B^i
\, 
\right)
T_{J,\B \R}
\right]^M
\right|_{0,0}
\\
& =
\left.
\frac{\GR T_{J,\R \B} \GB T_{J,\B \R}}
{1-\GR T_{J,\R \B} \GB T_{J,\B \R}}
\right|_{0,0},
\end{split}
\end{equation}
where
\begin{equation}
\GR = (1-T_\R)^{-1},
\label{eq:GR}
\end{equation}
and
\begin{equation}
\GB = \int \frac{d^3 \vec{k}}{(2 \pi)^3}\, \frac{T_\B(\vec{k})^2}{1-T_\B(\vec{k})}
\, .
\label{eq:GB}
\end{equation}

\section{Results}
\label{sec:Results}
In the grand canonical ensemble the average length of the polymer is
given by $\langle N \rangle = \partial_\mu\log(\Gamma) = z \partial_z \log(\Gamma)$.
We are interested in the limit of very long polymers, where $\langle N \rangle\to \infty$.
This limit is obtained for a specific choice of the fugacity $z=e^\mu$, such that:
\begin{enumerate}
\item{There are infinitely many repetitions of native (\R) and molten
(\B) segments.  This occurs for a value of $z$ such that the largest
eigenvalue of $[\GR T_{J,\R\B} \GB T_{J,\B\R}] (z)$ equals $1$.}
\item{The size of an individual bubble diverges. In this case the singularity
arises from $\partial_z \GB (z)$.}
\item{The hypothetical possibility of an infinitely long native (\R)
segment does not arise, as $\partial_z \GR (z)$ only diverges for
values of $z$ that already cause one of the previous two cases to
occur.}
\end{enumerate}

As in the Poland-Scheraga model~\cite{Poland66}, case (1) corresponds
to a partially melted double strand (mixed phase of R and B segments),
while case (2) corresponds to a fully melted state comprised of one
bubble (bubble phase).

\subsection{Phase Diagram}
A phase transition between the two phases occurs when there are both
infinitely many repetitions of \R~and \B~segments, {\it and} the
average bubble size diverges.  Since our model has 6 parameters, we
have to select appropriate subspaces for the display of phase diagrams.  We choose to
regard the bending energies $\Jr$, $\Jb$, $\Jj$, and the joint
Boltzmann weight $\wj=e^{\epsj}$ as parameters, and display the phase
diagrams as a function of the force $f$ and the dimensionless
energy $\epsr=\log(\wrod)$.  The latter may be regarded as a stand-in
for an inverse temperature, since it is related to an actual
energy after division by $k_BT$. 
(As discussed following Eq.~(\ref{eq:Erods}), negative values of $\epsr(f)$ do not pose a problem,
as the actual binding energy is $k_BT(\Jr+\epsr)$.) 

\begin{figure}[t]
\includegraphics[width=11cm]{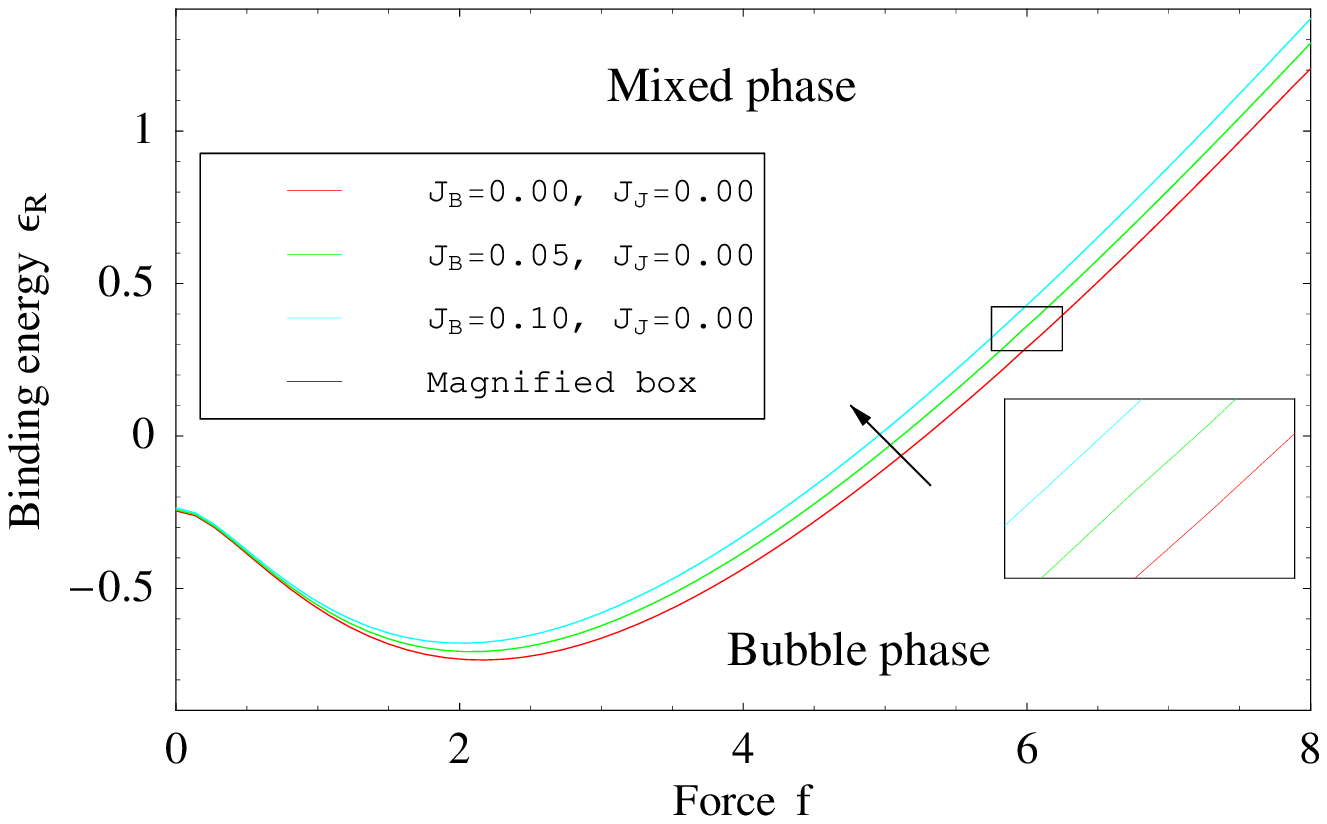}
\includegraphics[width=11cm]{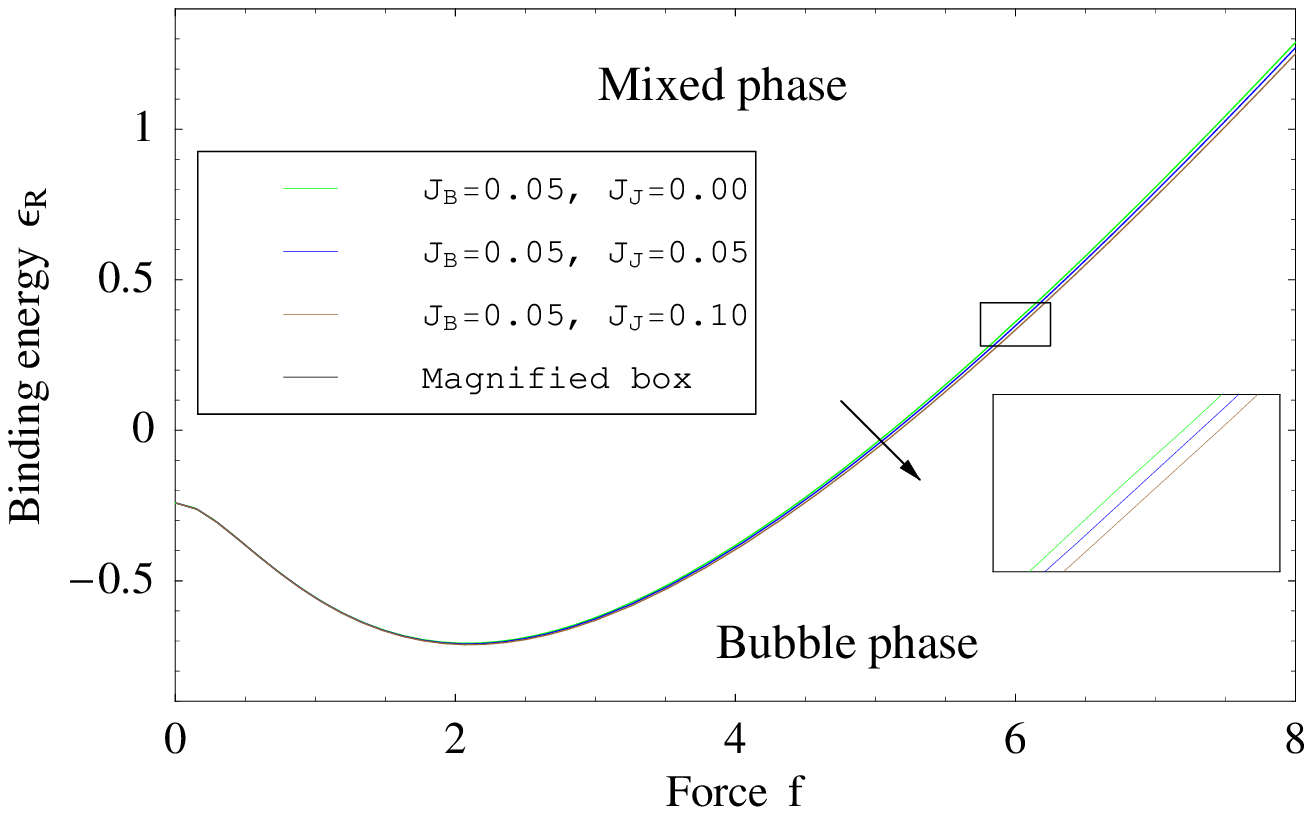}
\caption{The phase boundaries $\epsr(f)$ separate the partially melted
R-B phase (above the line) from the fully melted bubble phase (below
the line), for various combinations of $\Jb$ and $\Jj$.  Top: The
bubble stiffness $\Jb$ is increased.  Bottom: The joint stiffness
$\Jj$ is increased. In both plots, the arrow indicates the direction
in which the phase boundary moves.  In all curves
$|\vecr|=1.0,\,|\vecb|=1.7,\,\Jr=5.0$, and $\epsj=1.0$.}
\label{fig:PhaseTrans}
\end{figure}

Some typical phase boundaries  $\epsr(f)$ are presented
in Fig.~\ref{fig:PhaseTrans}.
We used the Mathematica software package on an Intel Pentium 3 GHz
desktop computer to obtain the phase diagrams, each of which took a
few hours of computational time when we included partial waves up to
$l=1$ in the bubble (B) partition functions, and $l=5$ in the native
(R) partition functions. It should be possible to reduce the
computational time significantly by using more appropriate
software. The truncation of the transfer matrices at these partial
waves was justified considering the small bending parameters chosen
for the bubbles and the joints.

Each solid curve depicts the phase boundary for a particular choice of
parameters.  All curves correspond to rather stiff \R-segments with
$\Jr=5.0$, but for different choices of the bending bending parameters
$\Jb$ and $\Jj$.  The upper portion of each figure corresponds to the
partially melted native state, which contains both \R~ and \B~
segments.  There is no native R segment left in the lower portion,
and  the polymer is a single bubble below the
phase transition line.   The lowest
curve in the top figure (red in the online version) corresponds to no
bending in the bubble or vertex; $\Jb=\Jj=0$. As we increase $\Jb$ to
larger values of 0.05 and 0.1 in the top figure (green and cyan,
online), the bubble phase becomes more stable. (The phase boundary
moves up as indicated by the arrow.)
If we now fix $\Jb=0.05$, and increase $\Jj$ to values of 0.05 and 0.1
in the bottom figure (blue and brown online) we find that $\epsr(f)$
moves down, as indicated by the arrow. The mixed phase is stabilized
by stiffening the joints between R- and B-segments.

Several features of these phase diagrams are now commented on in more detail.

\subsubsection{Explanation of the trends in phase diagrams}
With our choice of parameters, increasing the value of any stiffness parameter
$J$, makes the corresponding segment (or joint) more favorable.
This is because the corresponding Boltzmann weights are monotonically
increasing functions of $J$ (as are the modified spherical Bessel functions $i_\alpha(J)$).
For example, for a bubble segment the overall weight increases with $J_B$,
despite the fact that there are fewer configurations (and hence reduced) entropy
for the stiffened and stretched bubble. These trends are further magnified
at larger force as the stretched segments gain even more weight by
aligning to the force, as can be easily seen in Fig.~\ref{fig:PhaseTrans}.

\subsubsection{Reentrance \& the phase boundary at low and high forces}
An interesting feature of the phase boundary in
Fig.~\ref{fig:PhaseTrans} is its {\em reentrance}, namely for certain
choices of $\epsr$ ($-0.8<\epsr<-0.2$) the mixed R-B phase is
stable at intermediate values of the force, but melts at both weak and
strong force.  This reentrance is also present in another model of
denaturation which incorporates excluded volume effects in the
bubbles, but no bending rigidity~\cite{Hanke08}.

This feature can be explained by examining the
limiting behaviors of the phase boundary at small and large $f$.  
For large $f$, the polymer (whether in native or denatured state) is
stretched along the direction of the force. 
The contribution from  entropy is relatively small in this limit,
and one can estimate the location of the phase boundary by
comparison of energies: The energy of a fully stretched rod segment
is $\Jr+\epsr+f|\vecr|$ per base-pair. If the two strands are separated
the energy changes to $2\Jb+f|\vecb|$. The transition occurs
for $\epsr\approx 2\Jb-\Jr+(|\vecb|-|\vecr|)f$, which has a positive
slope since  \B~strands have longer monomers and are favored by the force.

As shown in Appendix~\ref{app:phasetransslope}, this argument
can be made more rigorous and extended to all cases where the contribution
from the joints can be ignored. In such cases the slope of the phase
boundary is given by $\partial_f \epsr
={\langle L_B \rangle}/{\langle N_B \rangle} - {\langle L_R
\rangle}/{\langle N_R \rangle}$, where $\frac{\langle L_X
\rangle}{\langle N_X \rangle}$ denotes the average length per monomer,
calculated for each segment type (\B~segment or \R~segment)
treated separately.

At zero force, the {\em average} end-to-end extension of each segment, 
$\langle L_X\rangle$, is zero by symmetry. 
The extension for small $f$ is linear, with a
force constant (susceptibility) that is easily related to the {\em
variance} of the end-to-end extension at $f=0$.  Since the change in
free energy is proportional to $f^2$, the phase boundary is also
quadratic in this limit.  In the absence of joint stiffness, the
curvature of the transition line can be related to the difference in
susceptibilities by $\partial_f^2 \epsr ={\langle L_B^2
\rangle_c}/{\langle N_B \rangle} - {\langle L_R^2
\rangle_c}/{\langle N_R \rangle}$, see Appendix~\ref{app:phasetransslope} 
for a derivation. Here, ${\langle L_X^2
\rangle_c}/{\langle N_X \rangle}$ denotes the variance in length of
rods or bubbles per monomer, computed for one rod segment or one
bubble segment subject to the same fugacity and force as for the whole
molecule.  Since it is easier to rotate and align the more rigid
\R~segments in the direction of the force, their gain from the force
is larger, and small force favors the native double-stranded phase.

The reentrance in the phase diagram of reference~\cite{Hanke08},
mentioned in the first paragraph of this subsection, can be explained
with these expressions as well. 
In this paper different behaviors are obtained as a function of a
parameter $A$, which determines how statistically favored joints are.
They observe a reentrant phase diagram for $A=0.01$ (disfavoring joints)
but not for $A=1$ (many joints).
The variance per monomer in the lengths of the bubbles is roughly
constant (${\langle L_B^2\rangle_c}\propto{\langle N_B \rangle}$ as in a random walk),
but for stiff rods the variance grows as the average size 
(${\langle L_R^2\rangle_c}\propto{\langle N_R \rangle^2}$ as in a directed walk).
For small $A$, there are few joints and longer rods just
after the phase transition into the mixed bubble--rod phase, making the
variance per monomer large. According to the above formula for
$\partial_f^2 \epsr$ this leads to reentrance. Note that excluded volume effects
do not substantially modify this argument.

Williams et al. experimentally observe (Figure~5 of Ref.~\cite{Williams02})
a reentrance in the phase boundary when they fit their
data to a simple model.  Unfortunately, the area of interest is merely
extrapolated and the transition is not probed at high enough temperatures
and low forces ($\approx 90^\circ$) to unambiguously verify reentrance.

\subsection{Force--Extension Isotherms}
\begin{figure}[t]
\includegraphics[width=\columnwidth]{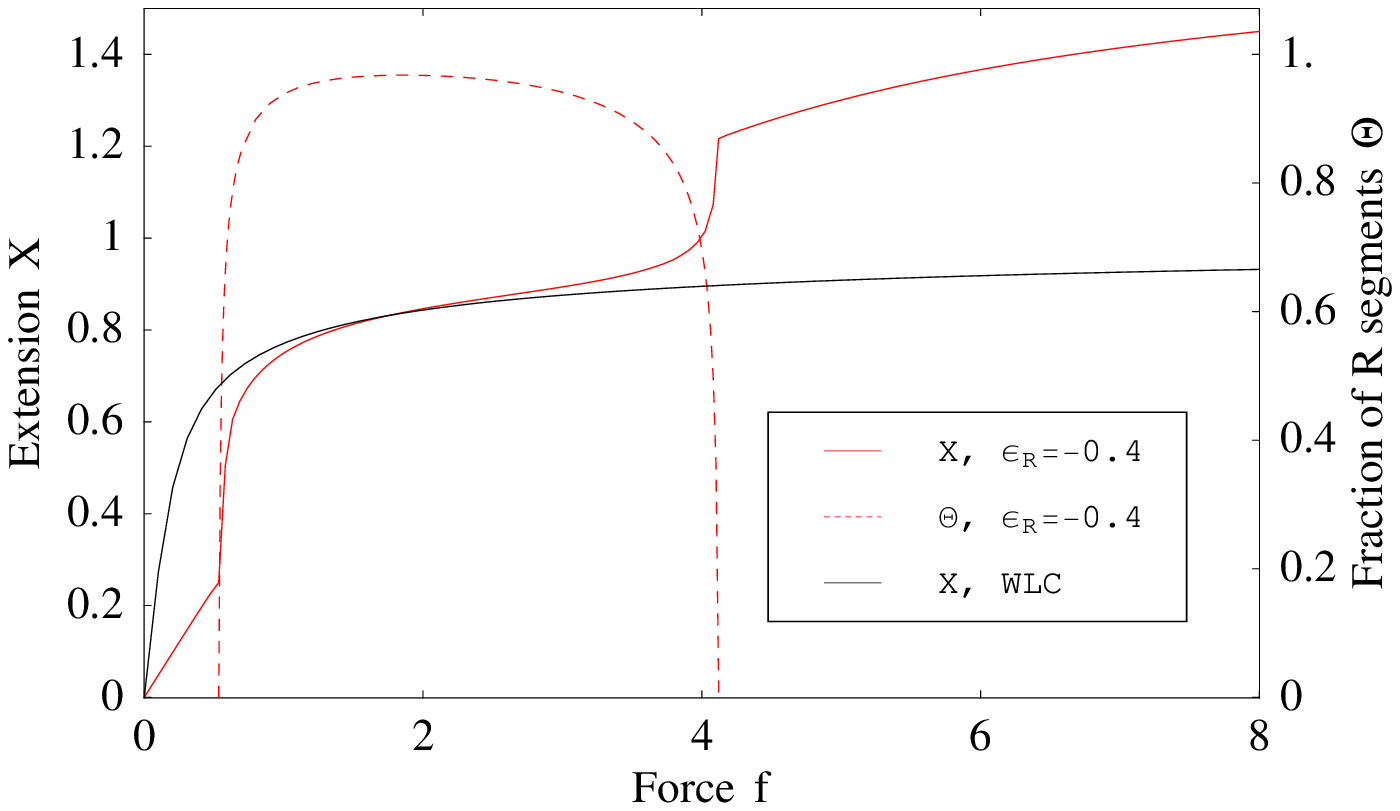}
\includegraphics[width=\columnwidth]{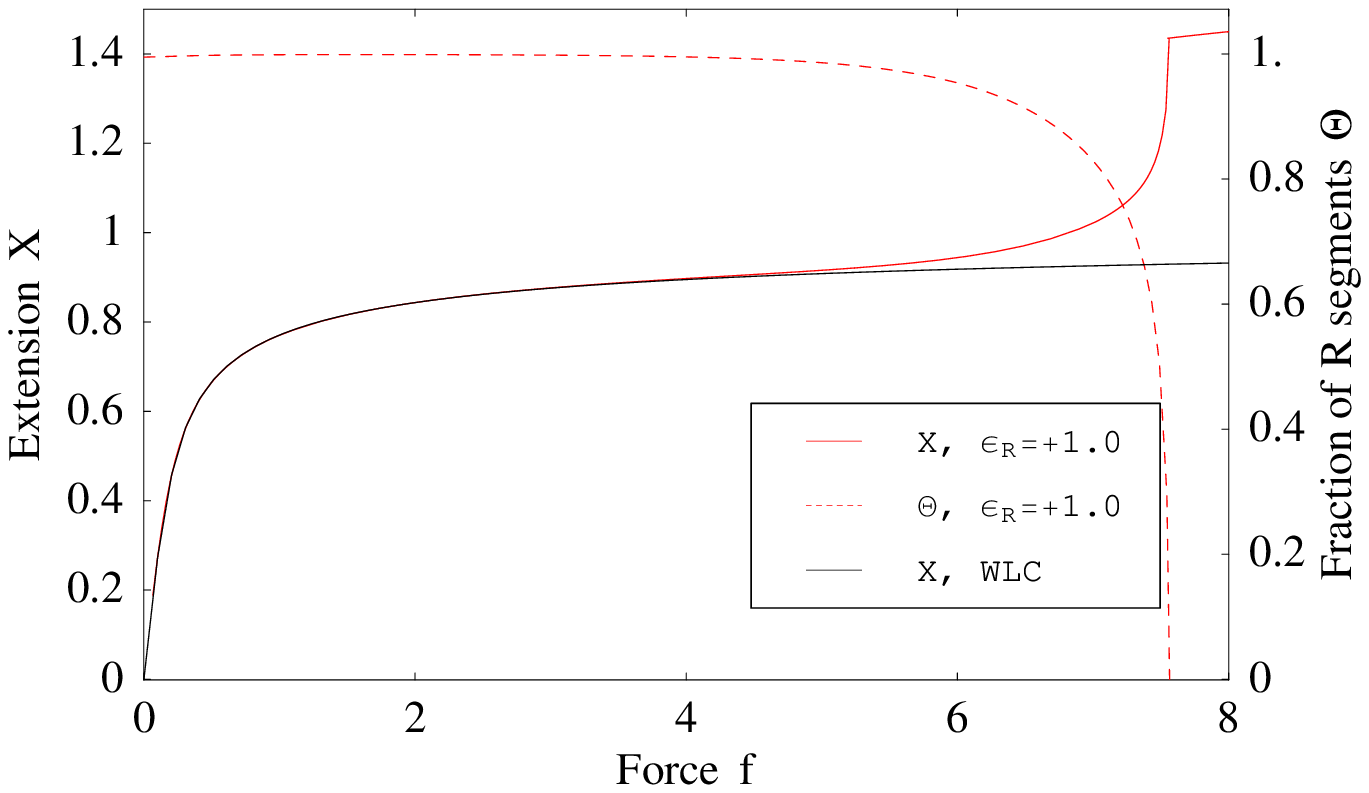}
\caption{ 
Comparison of the extension curves $X(f)=\langle L_z\rangle / \langle N |\vecr| \rangle$ (solid lines) for a double-stranded polymer, and a single stranded wormlike chain. The top panel corresponds to a case where melting is reentrant ($\epsr=-0.4$), while there is a single denaturation transition in the bottom panel for $\epsr=1.0$ (cf. Fig.~\ref{fig:PhaseTrans}). The fraction of native (double stranded) polymer, $\Theta\equiv\langle N_r\rangle/\langle N\rangle$, is indicated by dashed lines.
All curves correspond to $\Jb = \Jj = 0$, $|\vecr|=1.0,\,|\vecb|=1.7,\,\Jr=5.0$, and $\epsj=1.0$ for the double strands.}
\label{fig:ForceFraction}
\end{figure}
An important probe of phase behavior comes from the force--extension curves,
in which the end-to-end distance of the polymer is measured as a function
of increasing force.
Without loss of generality and to simplify calculations, these curves
are obtained for $\Jb=\Jj=0$ (with $\Jr=5.0,\,\epsj=1.0$).  For
comparison, we also plot the curves corresponding to the pure
worm-like chain (WLC) model in black (containing only \R~segments for
$\wj=0$).  The plotted `extension' is the average length along the
force direction, made intensive and dimensionless by dividing by the number of
monomers $\langle N\rangle$, and the monomer length $|\vecr|$ of the
\R~segment, i.e.
\begin{equation}
X=\frac{\langle L_z\rangle}{\langle N |\vecr| \rangle}
= \frac{\partial_f \Gamma}{\partial_\mu \Gamma}
= -\frac{d}{df}\left(\ln z\right)\big{|}_{\Gamma=\infty} .
\label{Fexten}
\end{equation}
The two panels in Fig.~\ref{fig:ForceFraction} were selected to correspond to parameters with (top) and without (bottom) reentrant melting.
(Consider horizontal lines in Fig.~\ref{fig:PhaseTrans} for $\epsr=-0.4$ and
$\epsr=1.0$, respectively.)
In both cases, the force-extension curves for the double-stranded polymer track the behavior of the worm-like chain closely in the mixed R-B phase, but deviate significantly in the denatured phases; most pronouncedly for the re-entrant transition.

\begin{figure}[t]
\includegraphics[width=\columnwidth]{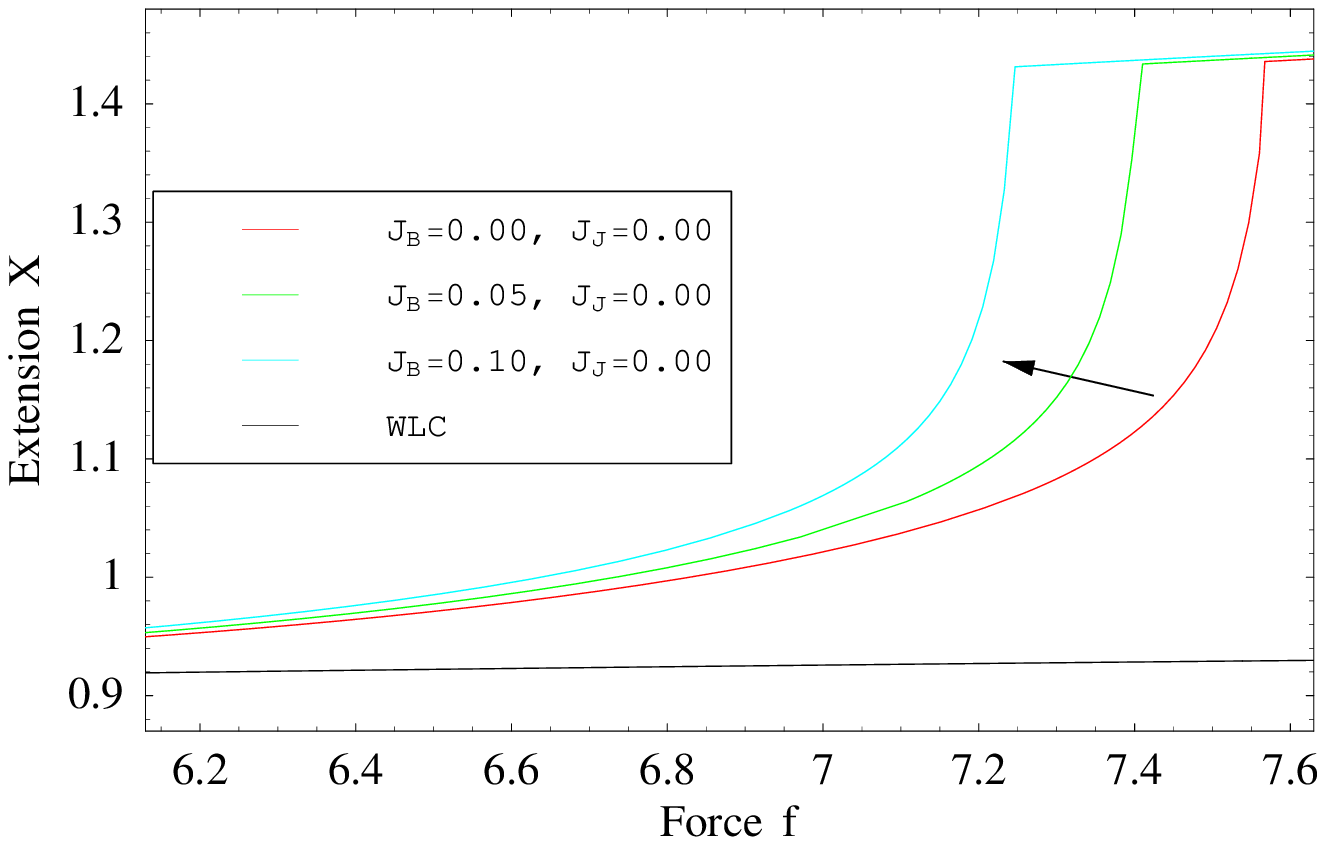}
\includegraphics[width=\columnwidth]{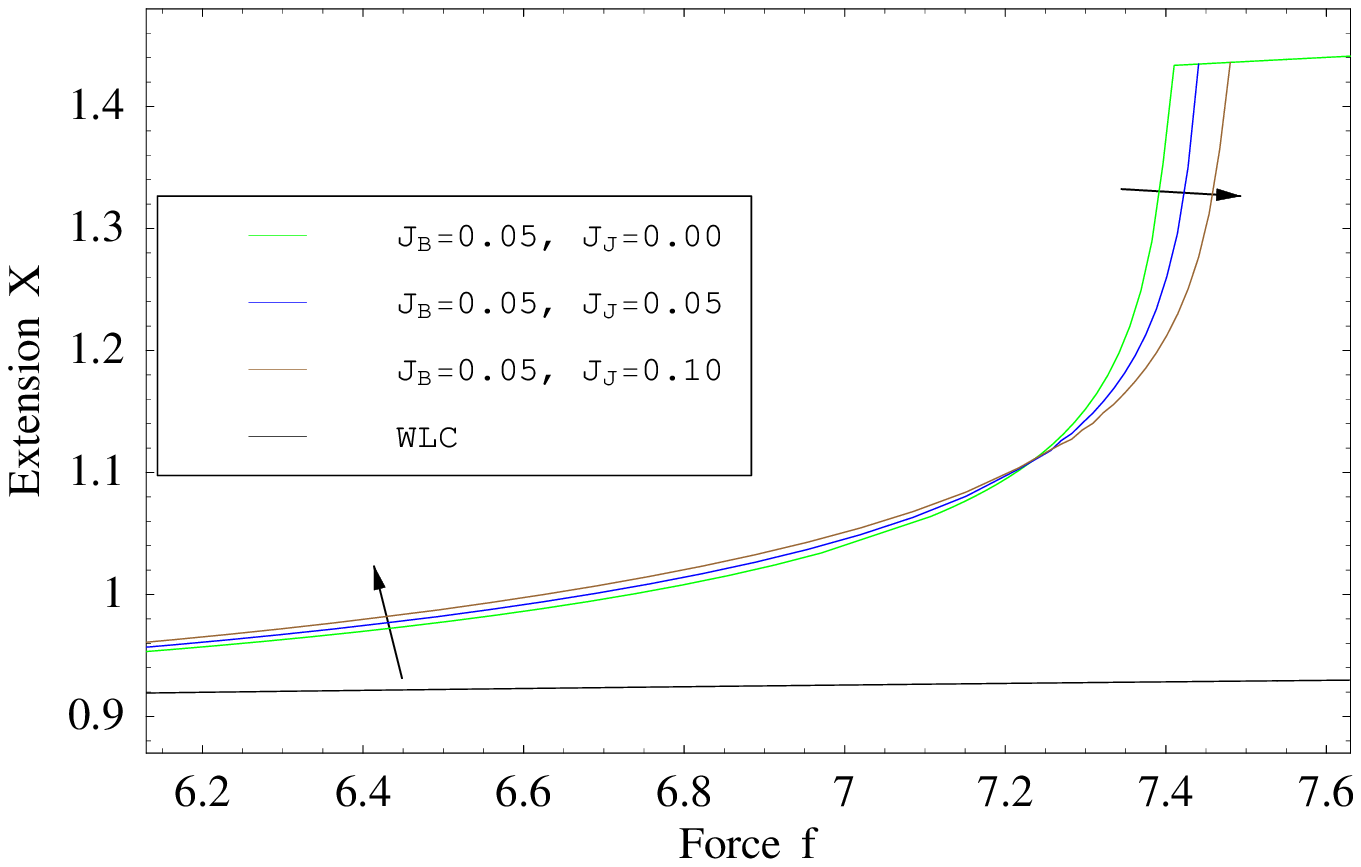}
\caption{Detailed view of the dimensionless extension 
$X=\langle L_z\rangle/\langle N |\vecr|  \rangle$ as a function of the force $f$, 
close to the phase transition, for various  combinations of $\Jb$ and $\Jj$.
    Top -- The bubble stiffness $\Jb$ is increased.
  Bottom -- The joint stiffness $\Jj$ is increased. In both plots, the arrows
  indicate the direction the extension curve moves.
  In all curves $|\vecr|=1.0,\,|\vecb|=1.7,\,\Jr=5.0$, and $\epsj=1.0$.
}
\label{fig:Fextendetail}
\end{figure}

As mentioned earlier, current experiments indicate that the WLC model
describes the extension of DNA accurately for small applied forces,
but fails at large forces due to the appearance of an over-stretched region.
In Fig.~\ref{fig:Fextendetail} we probe the corresponding region in more detail for our model, exploring the effect of bending rigidities (for $\epsr=1.0$ with a single transition).
The top panel depicts the effect of increasing the bubble stiffness $\Jb$, which makes the transition region appear sharper.
Increasing the joint stiffness $\Jj$ (bottom panel) has the opposite effect.
An interesting feature of the bottom panel is that the trends in $X(f)$ are not monotonic in $\Jj$, decreasing the extension for weaker force, and increasing it for larger force, leading to a crossing point in between.
The reader should note that we have taken $\epsj=1.0$ in all curves,
making the joints favorable and common. This choice is made  to
exaggerate the effect of the joint bending for display, as well as to
broaden the phase transition in Fig.~\ref{fig:Fextendetail}, thus highlighting the features of our model. A more realistic value,
namely $\epsj$ small or negative, gives qualitatively similar results.

\subsection{$\Theta$: Native (R) fraction}
Figure~\ref{fig:ForceFraction} also includes the native fraction
$\Theta$ as a function of $f$, depicted by the dashed curves.  This
is defined as the fractional amount of \R~segments in the polymer,
which can be computed from
\begin{equation}
\Theta \equiv \frac{\langle N_R \rangle}{\langle N \rangle} =
\frac{\wrod \partial_{\wrod} \Gamma}{z \partial_z \Gamma} .
\label{Theta}
\end{equation}
Note that $\Theta$ goes to zero continuously on approaching the bubble phase, underscoring the second order nature of the phase transition.  It approaches zero rapidly, but in a {\em linear} fashion. 
This is because the phase transition in our model belongs to the same 
universality class as the classic Poland-Scheraga model~\cite{Poland66}.  The addition of bending rigidity is irrelevant close to the phase transition, and excluded volume effects (which do modify the universality~\cite{Kafri02,Hanke08}) are not included in our model.

\section{Discussion}
\label{sec:Discussion}
We have introduced a formalism to address the role of bending rigidity
in the denaturation of double-stranded polymers, DNA providing a prime
example.  There has been some controversy on interpreting experimental
results for melting of DNA, or its denaturation by force.  There is
strong theoretical indication that the melting of a uniform
double-stranded polymer should be discontinuous due to excluded volume
effects~\cite{Kafri02}.  The discontinuity may
be masked in experiments because of the inherent
inhomogeneity of DNA~\cite{Rudnick07,Tang01}, or by finite-size
effects.  The rigidity of DNA should play an important role in the
latter, as longer persistent segments are less susceptible to
fluctuations and excluded-volume effects.  It is thus necessary that
comparison of models to experiment should include the effect of
rigidity, as we have attempted here. More generally, our formalism can
be extended to decribe the unravelling of any number of strands, for example from 1 to 3 in the case of collagen \cite{Thompson01,Gutsmann04}.

\begin{acknowledgments}
M.~P.~H. is supported in part by funds provided by the U.S.~Department of
Energy (D.O.E.) under cooperative research agreement DE-FC02-94ER40818.
S.~J.~R. and M.~K. are supported by NSF grant  DMR-04-26677. 
\end{acknowledgments}

\begin{appendix}

\section{Gaunt coefficients}
\label{app:Gaunt}
The Gaunt coefficients are defined as
\begin{equation}
C_{\alpha,\beta,\gamma}\equiv\int_{\mathbf{S^2}}
Y_\alpha(\hatr) Y_\beta(\hatr) Y_\gamma(\hatr) \, d^2\hatr\quad ,
\end{equation}
where $Y_\alpha$ is the spherical harmonic with indices $(l_\alpha,
m_\alpha)$. If a bar is put on top of an index of $C$, the
corresponding spherical harmonic in the integrand is complex
conjugated. The relation
\begin{equation}
Y_{l,m}^* = (-1)^m Y_{l,-m}\, ,
\end{equation}
can be used to relate a modified Gaunt coefficient with barred indices
to one without barred indices. A well-known expression for the Gaunt
coefficient in terms of Wigner $3j$-symbols is~\cite{Mathworld}
\begin{equation}
\begin{split}
C_{\alpha,\beta,\gamma} & = \sqrt{\frac{(2l_\alpha+1)(2l_\beta+1)(2l_\gamma+1)}{4\pi}} \\
& \times
\begin{pmatrix}
l_\alpha & l_\beta & l_\gamma \\
0 & 0 & 0 \\
\end{pmatrix}
\begin{pmatrix}
l_\alpha & l_\beta & l_\gamma \\
m_\alpha & m_\beta & m_\gamma \\
\end{pmatrix}.
\end{split}
\end{equation}
Using the properties of the Wigner $3j$-symbols one can restrict and
simplify the sums appearing in the partition function.

\section{Slope of the phase boundary}
\label{app:phasetransslope}
When the joint stiffness $\Jj$ vanishes, the partition function
matrices $\GR$ and $\GB$ in Eqs.~(\ref{eq:GR}) and (\ref{eq:GB}) reduce to
real-valued functions. As discussed in the Sec.~\ref{sec:Results}, two
conditions have to be met at the phase transition, $\GB\GR=1$ and
$\partial_z \GB=\infty$, where in the former equation all
multiplicative factors from the joints are  absorbed
in either of the two partition functions.  
Together, these two conditions set the value of $\mu=\log(z)$
and $\epsr=\log(w)$ {\em along the phase boundary}.
All manipulations below are then performed as the boundary point
is changed by varying the force $f$.

Noting that, the first condition is equivalent to 
\begin{equation}
\log \GB(\mu,f) + \log \GR(\mu,f,\epsr)=0\quad, 
\end{equation}
its variations are obtained, by taking one total derivative with respect to $f$, as
\begin{equation}
\begin{split}
0 = & \partial_f \log \GB + \partial_\mu \log \GB\, \partial_f \mu + \\
& \partial_f \log \GR + \partial_\mu \log \GR \, \partial_f \mu +
\partial_{\epsr} \log \GB\, \partial_f \epsr \\
= & \langle L_\B \rangle + \langle N_\B \rangle \partial_f \mu + \\
& \langle L_\R \rangle + \langle N_\R \rangle (\partial_f \mu + \partial_f \epsr).
\end{split}
\label{eq:oneder}
\end{equation}
From the second condition we get
\begin{equation}
\left. \partial_f \mu \right|_{\partial_\mu \GB=\infty} =
- \frac{\partial_f \partial_\mu \GB}{\partial_\mu \partial_\mu \GB} =
- \frac{\langle L_\B N_\B \rangle}{\langle N_\B^2 \rangle}.
\end{equation}
But because the condition $\partial_\mu \GB=\infty$ of infinite bubble
length is the same as $\partial_\mu \log \GB=\infty$, one can
equivalently express
\begin{equation}
\left. \partial_f \mu \right|_{\partial_\mu \log \GB=\infty} =
- \frac{\partial_f \partial_\mu \log \GB}{\partial_\mu \partial_\mu \log \GB} =
- \frac{\langle L_\B N_\B \rangle_c}{\langle N_\B^2 \rangle_c}\, ,
\end{equation}
where the subscript `c' indicates a cumulant in place of a moment.
From combining both expressions it follows that
\begin{equation}
\left. \partial_f \mu \right|_{\partial_\mu \GB=\infty} =
- \frac{\langle L_\B \rangle}{\langle N_\B \rangle}.
\label{eq:dermu}
\end{equation}
Plugging this result into Eq.~(\ref{eq:oneder}), one obtains
\begin{equation}
\partial_f \epsr = \frac{\langle L_B \rangle}{\langle N_B \rangle} -
\frac{\langle L_R \rangle}{\langle N_R \rangle}.
\end{equation}

Note that at zero force all averages with only one $L_\B$ or $L_\R$
are zero by symmetry. Taking another total derivative of
Eq.~(\ref{eq:oneder}), and dropping the terms that vanish for this
reason, one finds that at $f=0$
\begin{equation}
\begin{split}
0 = & \langle L_\B^2 \rangle_c + \langle N_\B \rangle \partial_f^2 \mu + \\
& \langle L_\R^2 \rangle_c + \langle N_\R \rangle (\partial_f^2 \mu + \partial_f^2 \epsr).
\end{split}
\label{eq:twoder}
\end{equation}

Taking another derivative of $\mu$ in Eq.~(\ref{eq:dermu}) one finds
(for $f=0$)
\begin{equation}
\left. \partial_f^2 \mu \right|_{\partial_\mu \GB=\infty} =
- \frac{\langle L_\B^2 \rangle}{\langle N_\B \rangle}.
\label{eq:twodermu}
\end{equation}
Combining Eqs.~(\ref{eq:twoder}) and (\ref{eq:twodermu}) the desired slope
at $f=0$ is obtained as
\begin{equation}
\partial_f^2 \epsr = \frac{\langle L_B^2
\rangle_c}{\langle N_B \rangle} - \frac{\langle L_R^2
\rangle_c}{\langle N_R \rangle}.
\end{equation}

\end{appendix}


\end{document}